\documentclass[12pt]{article}

\usepackage{graphicx}
\usepackage{amsmath}
\usepackage{amssymb}
\usepackage{caption} 
\usepackage[square,sort&compress,numbers]{natbib} 
\setcitestyle{citesep={,}}
\usepackage{color}
\usepackage{hyperref} 

\definecolor{darkgreen}{rgb}{0,.7,0}

\hypersetup{
    pdffitwindow=true,      
    pdftitle={Shock Treatment: Heavy Quark Drag in a Novel AdS Geometry},    
    pdfauthor={W. A. Horowitz and Yuri V. Kovchegov},     
    pdfsubject={AdS/CFT heavy quark drag},   
    pdfkeywords={AdS/CFT correspondence, heavy quark energy loss, jet quenching}, 
    pdfnewwindow=true,      
    bookmarksnumbered=true,
    colorlinks=true,       
    linkcolor=red,          
    citecolor=darkgreen,        
    filecolor=magenta,      
    urlcolor=cyan,           
    menucolor=blue,
    baseurl=http://www.arxiv.org/abs/,
}

\newcommand{\sectionref}[1]{Sect.~\ref{#1}}

\newcommand{\be}{\begin{equation}}
\newcommand{\ee}{\end{equation}}
\newcommand{\bfig}{\begin{figure}}
\newcommand{\efig}{\end{figure}}
\newcommand{\bea}{\begin{eqnarray}}
\newcommand{\eea}{\end{eqnarray}}

\newcommand{\pt}{$p_T$ }

\newcommand{\lowpt}{low-\pt}

\newcommand{\highpt}{high-\pt}

\newcommand{\thooft}{'t Hooft }

\newcommand{\infinity}{\infty}

\newcommand{\eq}[1]{Eq.~(\ref{#1})}

\newcommand{\fig}[1]{Fig.~\ref{#1}}

\begin{document}

\begin{center}
  {\bf \Large Shock Treatment: Heavy Quark Drag \\[5mm] in a Novel AdS Geometry} \\[1cm]
W.\ A.\ Horowitz,\footnote{horowitz@mps.ohio-state.edu} \ 
Yuri V.\ Kovchegov\footnote{yuri@mps.ohio-state.edu} \\[3mm]
  {\it\small Department of Physics, The Ohio State University, 191
    West Woodruff Avenue, \\ Columbus, OH 43210, USA}
\end{center}

\begin{abstract}
We calculate the drag force on a heavy quark hit by a shock wave, thus generalizing the strongly coupled AdS/CFT heavy quark drag calculations to both hot and cold nuclear matter.  The derivation employs the trailing string configuration, similar to that used in the literature for a quark moving through a thermal medium, though in the shock metric the string profile is described by a much simpler analytic function. Our expression for the drag depends on the typical transverse momentum scale of the matter in the shock. For a thermal medium this scale becomes proportional to the temperature, making our drag coefficient and momentum limit of applicability identical to those found previously. As the shock wave can be composed of either thermalized or non-thermalized media, our derivation extends the existing drag calculations to the case of arbitrarily distributed matter.
  ~
\end{abstract}


\section{Introduction}

The simultaneous emergence of the anti de-Sitter/conformal field
theory (AdS/CFT) correspondence
\cite{Maldacena:1997re,Gubser:1998bc,Witten:1998qj,Aharony:1999ti} as
a tool to study strongly coupled systems (see
\cite{Gubser:2009md,Gubser:2009sn} for reviews) and the failure of
perturbative QCD (pQCD) techniques \cite{Wicks:2005gt} to
quantitatively describe the observed heavy ion physics phenomena has
inspired both a critical reevaluation of the traditional methods and
an explosion of new research
\cite{Kovtun:2004de,Janik:2005zt,Liu:2006nn,Herzog:2006se,Gubser:2006bz,CasalderreySolana:2007qw,Horowitz:2007su}.
In particular a detailed theoretical understanding of \highpt
particles and jets, coupled with experimental measurements associated
with them, holds out the promise of uniquely probing the \highpt
physics, \lowpt collective physics, and their mutual interaction in a
bulk QCD medium.  From a theoretical and experimental perspective
heavy quark jets offer a rich new set of tools and challenges.
Previous AdS/CFT calculations
\cite{Liu:2006nn,Herzog:2006se,Gubser:2006bz,CasalderreySolana:2007qw,Horowitz:2007su,Mikhailov:2003er,Sin:2004yx,Chernicoff:2008sa,Kharzeev:2008qr,Chesler:2007sv}
of the heavy quark drag assumed a strongly coupled thermal medium of
${\mathcal N} =4$ super Yang-Mills (SYM) plasma. In this Letter we
extend the earlier work on the heavy quark drag in a thermal medium to
that of an arbitrary medium; in particular our formalism applies to
both hot and cold strongly coupled nuclear matter.

The AdS/CFT conjecture postulates a duality between certain field
theories and the compactification of Type IIB string theory in various
geometries
\cite{Maldacena:1997re,Gubser:1998bc,Witten:1998qj,Aharony:1999ti}.
In particular $\mathcal{N}=4$ SYM theory is dual to Type IIB string
theory in AdS$_5\times$S$^5$.  What makes the conjecture so useful,
but also so difficult to prove, is that the weak coupling limit of one
theory is dual to the strong coupling limit of the other.  Of
especial interest to the heavy ion community is the strong coupling
limit of QCD, where the only previous theoretical tool was numerical
lattice simulations. While lattice simulations remain the only method
to obtain quantitative results for strongly-coupled QCD, one may argue
that for some observables similarities between QCD and $\mathcal N =
4$ SYM theory can be exploited to improve our qualitative
understanding of the former by performing calculations in the
latter. In the limit of large \thooft coupling and number of colors
$N_c$, the $\mathcal N = 4$ SYM theory is dual to an (often)
analytically tractable theory, the classical limit of string theory:
classical supergravity (SUGRA).

Despite the many differences between QCD and $\mathcal{N}=4$ SYM,
there have been a number of qualitative successes in applying the
AdS/CFT ideas to heavy ion phenomenology at RHIC.  Specifically, the
lattice result of a surprising deviation of the entropy density above
the deconfinement temperature from the Stefan-Boltzmann limit
\cite{Karsch:2000ps,Cheng:2007jq} is easily understood analytically in
AdS/CFT \cite{Gubser:1996de,Klebanov:2000me}; the observed near
perfect fluidity of the QGP
\cite{Teaney:1999gr,Kolb:2000sd,Huovinen:2001cy,Teaney:2003kp,Song:2007ux}
comes from a shear viscosity to entropy ratio an order of magnitude
smaller than perturbative estimates, but in line with the AdS/CFT
calculations \cite{Kovtun:2004de}; the shockingly large suppression of
heavy quarks
\cite{Adler:2005xv,Abelev:2006db,Adare:2006nq,Mischke:2008qj,Morino:2008nc}
appears to be easier to describe in the AdS/CFT framework
\cite{Liu:2006nn,Herzog:2006se,Gubser:2006bz,CasalderreySolana:2007qw,Mikhailov:2003er,Sin:2004yx,Chernicoff:2008sa,Kharzeev:2008qr};
and the emergence of an away-side double hump structure in two
\cite{Adams:2005ph,Adler:2005ee,Adare:2008cqb} and three particle
\cite{Ulery:2005cc,Zhang:2007zzp} correlators, while suggestive of a
Mach cone formed by a supersonic jet
\cite{Stoecker:2004qu,CasalderreySolana:2004qm}, can currently only be
understood as the result of the near field energy and momentum
deposition derived using AdS/CFT
\cite{Friess:2006fk,Chesler:2007sv,Betz:2008wy}.

On the other hand, from both experimental and theoretical standpoints,
perturbative methods as applied to jet quenching physics should be
viewed with increased skepticism.  Early quantitative success in
describing the energy loss of light partons \cite{Baier:1996kr,Gyulassy:2000er,Vitev:2002pf} as
evidenced by the suppression of pions \cite{Adare:2008qa,Lin:2008zzi},
etas \cite{Adler:2006bv}, and the null control of direct gammas
\cite{Adler:2005ig,Miki:2008zz} has given way to an inability
\cite{Djordjevic:2005db,Wicks:2005gt,Armesto:2005mz} to simultaneously
describe even any two of the following four observables: the (1) pion
quenching and (2) similarly large suppression of non-photonic
electrons (NPE), the decay fragments of heavy quarks, and the
surprisingly large azimuthal anisotropy of the (3) hadrons from light
parton jets \cite{Adler:2006bw,Abelev:2008ed} and of the (4) NPE
\cite{Adare:2006nq}.  From a self-inconsistency standpoint the
assumption of a small coupling, $g \ll 1$, in \highpt energy loss has
never been true---at RHIC energies and temperatures $g\sim2$---with
the momentum scale, set by either the Gyulassy-Wang model
\cite{Gyulassy:1993hr} of Yukawa-like scattering centers or the
saturation scale, on the order of $0.5-1$
GeV$\sim\mathcal{O}(1)\Lambda_{QCD}$, far from perturbative.

An understanding of the limits of applicability, of the inherent
theoretical error, are required in order to either gain confidence in
or falsify a calculation with experimental data.  In deep inelastic
scattering (DIS) and Drell-Yan production, the rigorous framework of
factorization allows for a well-controlled pQCD expansion (see
\cite{Collins:1992xw} and references therein). For heavy ion
collisions it is not clear that factorization holds any longer
\cite{Collins:2007nk}.  Even supposing it does, the numerical value of
the momentum characterizing the scale at which pQCD methods break down
is not quantitatively known.  The AdS calculations are under even less
control.  Until a dual string geometry is found for QCD one cannot
quantitatively estimate the changes in observables from using a
different theory; there is no path through dual theories approaching
QCD whose parameterization may be used as an expansion parameter.

What one can do is explore the results from different theories and
geometries in an attempt to discover universal behavior. This may give
one reason to believe that the result will hold for QCD, at least at
the qualitative level.  In this Letter we follow this strategy by
considering the heavy quark momentum loss in a novel geometry.
Previous drag force calculations focused on motion of the gauge theory
quark in vacuum (empty AdS metric)
\cite{Mikhailov:2003er,Chernicoff:2008sa,Kharzeev:2008qr}, or in a
static thermal medium (black hole and black hole-like metrics)
\cite{Liu:2006nn,Herzog:2006se,Gubser:2006bz,CasalderreySolana:2007qw,Sin:2004yx,Chernicoff:2008sa}.
Here we examine the drag force on a static quark inside a shock wave.
In particular the shock wave can be a model of the dense medium
produced in a heavy ion collision. Equivalently it can also be a model
of a nucleus incident on a probe quark, giving rise to cold-matter
energy loss as is often considered in proton-nucleus collisions.

The paper is structured as follows. In \sectionref{sec:metric} we
present the shock wave metric and solve the string equations of motion
to find the trailing string solution, similar to 
\cite{Herzog:2006se,Gubser:2006bz,CasalderreySolana:2007qw}. We note
that the analytic form of the trailing string solution, given below in
\eq{zcubedsoln}, is considerably simpler than that found in
\cite{Herzog:2006se,Gubser:2006bz,CasalderreySolana:2007qw}. In
\sectionref{sec:daforce} we use the solution from \eq{zcubedsoln}
obtained in \sectionref{sec:metric} to calculate the drag force on the
quark inside the shock wave. Boosting the resulting expression to the
rest frame of the medium and taking its typical momentum to be set by
the temperature we obtain the same equation for the drag force as
derived in the original works
\cite{Herzog:2006se,Gubser:2006bz,CasalderreySolana:2007qw}. 
Just as the form of perturbative jet energy loss is independent of the
state of the medium, it turns out that the heavy quark energy loss in
a shock wave is exactly the same as that in a thermalized medium. Our
result is derived for a shock wave made of any distribution of matter:
due to Lorentz time-dilation the shock wave may represent a
``snapshot'' of any non-equilibrium matter. Hence our result
generalizes the drag force of
\cite{Herzog:2006se,Gubser:2006bz,CasalderreySolana:2007qw} to the
case of a non-thermal medium. We also study the limitations of our
calculation in \sectionref{sec:limits} and find that the momentum
``speed limit'' of applicability for the shock wave calculation is
parametrically exactly the same as for the black hole metric. We
conclude in \sectionref{sec:conc} by restating our main results and
discussing the relation between the quark mass in vacuum and in medium
for strongly-coupled gauge theories.


\section{Shock Metric, EOM, and the Trailing String Solution}
\label{sec:metric}

We will consider the generalized ``shock'' metric \cite{Janik:2005zt}
\begin{subequations}
\begin{align}
\label{lcmetric}
ds^2 & \equiv \, G_{\mu\nu} \, d x^\mu \, d x^\nu \, = \,
\frac{L^2}{z^2} \, \left[ -2dx^+dx^- + 2 \, \mu \, z^4 \, \theta(x^-)
  \, dx^{-2} + dx_\perp^2+dz^2 \right] \\
\label{metric}
& = \frac{L^2}{z^2} \, \left[ -\left(1- \mu \, z^4 \, \theta(x^-)
  \right) \, dt^2 - 2 \, \mu \, z^4 \, \theta(x^-) \, dt \, dx +
  \left(1+\mu \, z^4 \, \theta(x^-)\right)dx^{2}+dx_\perp^2+dz^2
\right],
\end{align}
\end{subequations}
where we have used the $x^\pm=(t \pm x)/\sqrt{2}$ normalization of
light-cone coordinates with $x = x^3$ and dropped the $d\Omega_5^2$
standard metric of the five-sphere in AdS$_5\times S^5$. As usual $L$
is the radius of the $S^5$ space. Also $dx_\perp^2 = (d x^1)^2 + (d
x^2)^2$ is the transverse part of the metric, and $\mu,\nu = 0, \ldots
, 4$.

The physical justification for describing \eq{lcmetric} as a shock
metric comes from the application of holographic renormalization
\cite{deHaro:2000xn}, which relates the metric in Fefferman-Graham
form \cite{Fefferman} in $d+1$ dimensions to the energy-momentum
tensor in the $d$ dimensional boundary theory.  For our particular
choice of metric, \eq{lcmetric}, the above prescription yields 
\begin{align}
  \label{tmu}
\langle T_{--} \rangle = \frac{N_c^2}{\pi^2} \, \mu \, \theta(x^-),
\end{align}
which describes an ultrarelativistic shock wave (shock front) moving
in the positive $x^3 = x$ direction.

\begin{figure}[!htb]
\centering
\includegraphics[width=14cm]{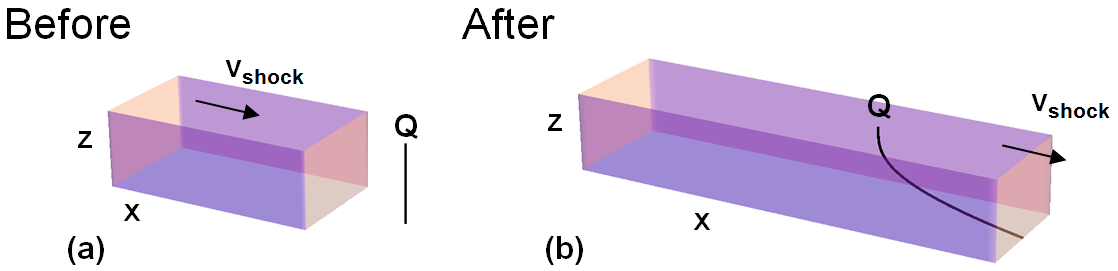}
\caption{\label{setup}
  Cartoon of the shock medium colliding with the heavy quark $Q$ in
  its rest frame. (a) Before the collision the string representing the
  heavy quark hangs from the D7 brane straight down to the D3 branes
  at the origin. (b) After the collision the string trails behind $Q$.
  As the shock is moving in the positive direction ($v_\mathrm{sh} >
  0$) in the heavy quark rest frame momentum actually flows up the
  string as the shock tries to transfer positive momentum to the
  quark; see \sectionref{sec:daforce} for more details.  }
\end{figure}

Previous calculations used $\mu \, \delta(x^-)$ as the coefficient of
the $dx^{-2}$ term in \eq{lcmetric} to represent the
Lorentz-contracted pancake of nuclear matter probed in DIS
\cite{Albacete:2008ze} or seen in heavy ion collisions
\cite{Albacete:2008vs}. Note that in those papers $\mu$ is a slightly
different quantity with units GeV$^3$, unlike GeV$^4$ here.  As
observed in \cite{Janik:2005zt} the coefficient of $dx^{-2}$ can be
any function of $x^-$; the resulting metric still satisfies Einstein's
equations. We take it $\propto\theta(x^-)$ to represent an incoming
dense medium of nuclear matter colliding with a heavy quark in its
rest frame. The setup is shown in \fig{setup},
where the shaded region is the incoming shock wave. The fifth 
dimension $z$ increases as one moves down the $z$ axis: the top of 
the diagram is the boundary of AdS$_5$ space ($z=0$); at the bottom ($z=\infinity$) 
is the stack of $N_c$ D3 color branes.  A finite mass fundamental
representation heavy quark in the 4D field theory corresponds in the
10 dimensional supergravity theory to the endpoint of an open
Nambu-Goto string terminating on a D7 brane; the other end of the open
string ends on the $N_c$ stack of D3 branes at $z=\infinity$.  The D7
brane wraps an $S^3\subset S^5$ and fills the asymptotically AdS space
from $z=0$ down to $z=z_M$ \cite{Karch:2002sh}.  Before the collision the string 
hangs straight down (left panel of \fig{setup}), while after the
collision the string trails behind the quark (right panel of
\fig{setup}).

Ordinarily the spatial direction picked out when using light cone
coordinates represents the beam direction. That would be the case
here if we apply our model to the description of energy loss in
proton-nucleus collisions, i.e., in cold nuclear matter.  For jet
energy loss in heavy ion collisions $x$ corresponds to the direction
of motion of the heavy quark in the lab frame, an orientation often
taken transverse to the beam.

While \cite{Herzog:2006se} extended the heavy quark drag calculations
to more general metrics of the black hole type, a novel feature of the
$G_{\mu\nu}$ used here is its lack of an event horizon.  If light can
pass through a surface coming from inside the suspected black hole,
the surface is not a true horizon.  For our metric a light ray moving
towards the boundary of AdS space both in the $z$ and $x$ direction
can cross the suspected horizon at $z_h = \mu^{-1/4}$.  Since the
light ray can therefore escape from inside the suspected black hole,
the surface at $z_h = \mu^{-1/4}$ is not a true horizon.

We are interested in the motion of a heavy quark in the background
specified by the metric of \eq{metric}.  The test string
action is \be
\label{ngaction}
\begin{split}
  S_{NG} & = -T_0\int d\tau d\sigma\sqrt{- g}, \\
  g \, & = \, \mbox{det} \, g_{ab}, \ \ \ g_{ab} = G_{\mu\nu} \,
  \partial_a X^\mu \, \partial_b X^\nu,
\end{split}
\ee 
where $G_{\mu\nu}$ is the spacetime metric of \eq{metric}, Greek
indices refer to spacetime coordinates, and Latin indices to
worldsheet coordinates.  $X^\mu=X^\mu(\sigma)$ specifies the mapping
from the string worldsheet coordinates $\sigma^a$ to spacetime
coordinates $x^\mu$.  The backreaction of the fundamental string
($\mathcal{O}(N_c)$) is neglected as compared to the
$\mathcal{O}(N_c^2)$ contributions from the adjoint fields of
$\mathcal{N}=4$ SYM \cite{Herzog:2006gh}.

Varying the action, \eq{ngaction}, yields the equations of motion:
\be
\label{eomone}
\begin{array}{lr}
\nabla_a P^a_\mu = 0, & P^a_\mu = \pi^a_\mu/\sqrt{-g} = -T_0 \, G_{\mu\nu} \, 
\partial^a X^\nu,
\end{array}
\ee 
where $\nabla_a$ is the covariant derivative with respect to the
induced metric, $g_{ab}$, and the $\pi^a_\mu$ are the canonical
momenta, \be
\label{canonical}
\pi^a_\mu = -T_0 \, \frac{\partial\sqrt{-g}}{\partial (\partial_a X^\mu)}.
\ee  

We limit our attention to only the nontrivial directions of $t,z,$ and
$x$, where $x$ is parallel to the direction of the heavy quark
propagation in the lab frame (although we will carry out our
derivations in the rest frame of the heavy quark).  $X^\mu(\sigma)$
maps into the $(t,x,z)$ coordinates; choosing the static gauge,
$\sigma^a=(t,z)$, the string embedding is described by a single
function, $x(t,z)$:  
\begin{align}
  X^\mu (\sigma) \, = \, \left( t, 0, 0, x(t,z), z \right).
\end{align}
The equations of motion for $x(t,z)$ can be derived in two equivalent
ways: plug in the $x(t,z)$ ansatz into \eq{eomone}; or first
substitute the ansatz into \eq{ngaction}, then vary the action.
Following the latter approach and denoting $\partial_t x=\dot{x}$ and
$\partial_z x=x'$ we find
\begin{gather}
\label{rootg}
-g = \frac{L^4}{z^4}\big(1+x'^2-\dot{x}^2-\mu z^4 (1-\dot{x})^2\big), \\
\label{eomtwo}
\frac{\partial}{\partial t} \left( \frac{\mu z^4 - (1+\mu
    z^4)\dot{x}}{z^4\sqrt{-g}} \right) + \frac{\partial}{\partial z}
\left( \frac{x'}{z^4 \sqrt{-g}} \right) = 0.
\end{gather}

One may determine the mass of the equivalent point particle in the 4D
field theory by examining the string motion outside the shock.  For
$\mu=0$, the constant velocity string $x(t,z)=x_0+vt$ is clearly a
solution to the equations of motion, \eq{eomtwo}; the action,
\eq{ngaction}, becomes 
\be S=-T_0 L^2\int_{z_M}^\infinity
dz\frac{1}{z^2}\int dt \sqrt{1-v^2}, 
\ee 
the action of a point particle of mass $M_q=T_0 L^2/z_M$. Hence we
have that 
\be\label{zmass}
z_M=\frac{T_0 L^2}{M_q} = \frac{\sqrt{\lambda}}{2\pi M_q},
\ee
where we have used the AdS dictionary to relate $T_0 L^2 = \sqrt{\lambda}/2\pi$.

\begin{figure}[!htb]
\centering
\includegraphics[width=12cm]{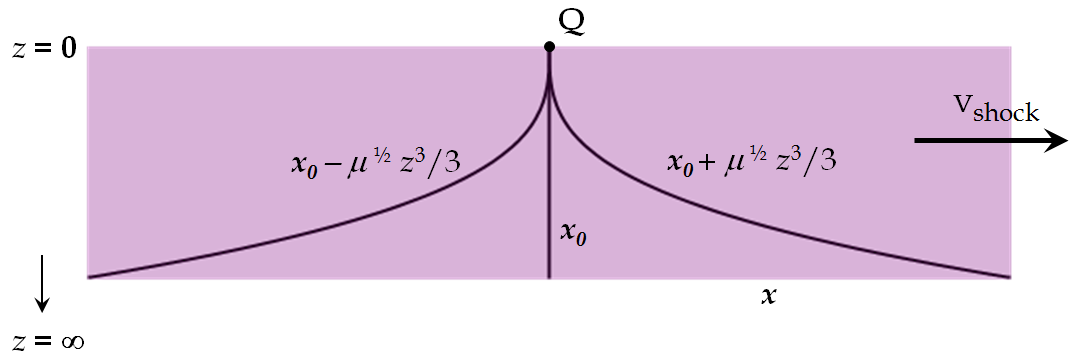}
\caption{
  The three solutions, $x(t,z)=\xi(z)=x_0, \,x_0 \pm\sqrt{\mu}z^3/3$,
  to the equation of motion for a static string inside the shock,
  \eq{eomthree}.  The constant solution ($x(t,z)=x_0$) is unstable, $S
  \sim i\infinity$.  We also throw away the unphysical time-reversed
  solution ($x(t,z)= x_0 - \sqrt{\mu}z^3/3$) by noting that any
  fluctuation would cause it to move toward the remaining (physical)
  one (see the text for details).  }
\label{solns}
\end{figure}

Assuming an asymptotic static solution, $x(t,z)=\xi(z)$, in a metric
for which the shock occupies all space and time, the induced metric
becomes time-independent.  The equations of motion, \eq{eomtwo},
reduce to
\begin{gather}
\label{eomthree}
\frac{\partial}{\partial z} \left(\frac{\xi'}{z^4 \sqrt{-g}} \right) = 0.
\end{gather}

Consider the first integral of the motion resulting from
\eq{eomthree}.  If we set the quantity in parentheses in
\eq{eomthree} equal to a constant $C$, then we can solve for $\xi'$:
\be
\label{xiprime}
\xi'(z) = \pm C z^2 \sqrt{\frac{1-\mu z^4}{1-C^2 z^4}}.  \ee There are
two cases to consider for a string hanging from the boundary down to
the stack of D3 branes: $C=0$ and $C\ne0$. (For other solutions of the
equations of motion see Appendix \ref{sec:othersolns}.) For $C\ne0$,
the constant of integration is fixed by considering the signs of the
numerator and denominator inside the radical as a function of $z$: for
small $z$ both are positive; for large $z$ both are negative.  In
order to avoid imaginary solutions, the numerator and denominator must
change signs at precisely the same value of $z$; thus $C=\sqrt{\mu}$.
This leads immediately to
\begin{align}
  \label{zcubedsoln}
  \xi(z) = x_0\pm\frac{\sqrt{\mu}}{3}z^3,
\end{align}
which are shown in \fig{solns}.  It is amusing to note that the
near-boundary expansion of the static quark solution for the black
hole metric (with horizon at $z=z_h$) is $x(t,z)\approx x_0+vt\pm
z^3/(3z_h^2)$.  However we would like to point out that the string
configuration in \eq{zcubedsoln} is valid for all $z$ in the shock
metric and is thus significantly simpler than that found in the black
hole metric. As we will see below, while being analytically simpler,
our solution retains all the main qualitative features of the black
hole case.

The sign ambiguity, resulting from the time-reversal symmetry of the
problem, can be fixed by taking the positive, trailing solution; the
negative sign has the string ``trailing'' in front of the heavy quark.
We find from \sectionref{sec:daforce} that, in the lab frame, the
physical solution loses momentum to the medium whereas the unphysical
solution gains it.  Additionally we will find in
\sectionref{sec:limits} that any fluctuation of the negative sign
string solution will lead to it approaching its time-reversed,
physical solution.

Since the heavy quark is static inserting \eq{zcubedsoln} back into
the action and integrating out the $z$ coordinate yields a point
particle action $S = -M_q \, \int dt$ with the vacuum heavy quark mass
$M_q$.  This should be contrasted to the result found in the black
hole metric for which the point particle mass is modified from that in
vacuum --- interestingly, it is decreased in medium.  We will discuss
this further in the conclusions, \sectionref{sec:conc}.

For $C=0$, $\xi=x_0$ is immediately found: the string hangs straight
down.  However plugging the solution back into the action gives 
\begin{align}
  \label{ststact}
  S = -T_0 \int dt \int_{z_M}^\infinity dz \frac{1}{z^2}\sqrt{1-\mu
    z^4}.
\end{align}
One sees that the IR (large-$z$) part of the $z$ integration gives
the action an infinite imaginary part.  We interpret this as an
infinitely unstable state that would immediately decay into the
physical trailing string solution.  This and the two solutions from
\eq{zcubedsoln} are shown in \fig{solns}.


\section{Drag Force}
\label{sec:daforce}

The drag force on the heavy quark in the SYM theory corresponds to the
momentum flow from the direction of heavy quark propagation down the
string, i.e., \ $dp/dt = -\pi^1_x$.  $\pi^a_\mu$ are the canonical
momenta and may be found from Eqs. (\ref{canonical}) and
(\ref{rootg}):
\begin{align}
\left( \begin{array}{c}
\pi^0_t \\
\pi^0_x \\
\pi^0_z \end{array} \right) & = \frac{T_0 L^4}{z^4 \sqrt{-g}}
\left( \begin{array}{c}
-1-x'^2+\mu z^4(1-\dot{x}) \\
\dot{x}-\mu z^4(1-\dot{x}) \\
-x'\big( \dot{x} - \mu z^4 (1 - \dot{x}) \big)
\end{array}
\right) \label{pi0} \\
\left( \begin{array}{c}
\pi^1_t \\
\pi^1_x \\
\pi^1_z \end{array} \right) & = \frac{T_0 L^4}{z^4 \sqrt{-g}}
\left( \begin{array}{c}
\dot{x}x'\\
-x'\\
-1+\dot{x}^2+\mu z^4(1-\dot{x})^2\\
\end{array}
\right) \label{pi1}
\end{align}
The ``momentum change'' of our heavy quark solution given by
\eq{zcubedsoln}, where momentum change is in quotation marks as the
quark is held static, is then
\begin{align}
  \label{momgain}
  \frac{dp}{dt} = - \pi^1_x = \frac{\sqrt{\lambda}}{2\pi}\sqrt{\mu}.
\end{align}

Since all our calculations have been in the heavy quark rest frame,
\eq{momgain} does not describe the momentum loss for a heavy quark in
the lab frame, the rest frame of the shock.  Even though formally the
shock propagates at the speed of light we will think of it as an
approximation to the physical setup of a medium moving almost on the
light-cone; then this rest frame is well defined.  Moreover we can now
also relate $\mu$ to properties of the medium by comparing the
energy-momentum tensor found from the metric using the holographic
renormalization procedure, \eq{tmu}, to one derived from the medium
properties.

\begin{figure}[!htb]
\centering
\includegraphics[width=\columnwidth]{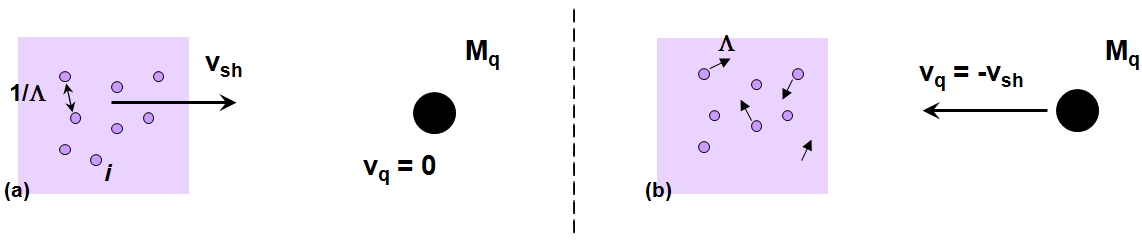}
\caption{\label{frames}
  Illustration of the shock wave colliding with a heavy quark in the
  (a) heavy quark rest frame and (b) medium rest frame.  In (a) the
  shock moves with velocity $v_{sh}$ in the positive $x$ direction; in
  (b) the heavy quark has momentum $p' \simeq \gamma M_q$ in the
  negative $x$ direction.}
\end{figure}

Following \cite{Albacete:2008ze}, we assume the medium is made up 
of $N_c^2$ valence gluons of the $\mathcal{N}=4$ SYM fields; 
see \fig{frames}.  If in the rest frame of the medium the particles 
are isotropically distributed with a typical momentum of order $\Lambda$---with 
associated inter-particle spacing of order $1/\Lambda$---then the 00 component of the 
stress-energy tensor in the rest frame of the shock is
\begin{align}
  \langle T'_{00} \rangle \, \propto \, N_c^2 \Lambda^4, 
\end{align}
where primes denote quantities in the rest frame of the medium and
proportionality is up to a constant numerical factor.  Changing into lightcone
coordinates and boosting into the rest frame of the heavy quark yields
\begin{align}\label{T--}
  \langle T_{--} \rangle = N_c^2 \Lambda^4 \gamma^2 = N_c^2 \Lambda^4
  \left(\frac{p'}{M}\right)^2,
\end{align}
where we assumed ultrarelativistic motion for the heavy quark in the
medium rest frame, $p' \simeq M\gamma$. (\eq{T--} can be viewed as a
definition of the scale $\Lambda$ absorbing numerical constants
arising from the precise definition of $\langle T'_{00} \rangle$ and
from the change to $\langle T_{--} \rangle$.) Comparing this with
\eq{tmu} we read off
\begin{align}
  \label{mu} \mu = \pi^2 \Lambda^4 \left(\frac{p'}{M}\right)^2.
\end{align}
We now have the right hand side of \eq{momgain} in terms of the
momentum of the heavy quark in the rest frame of the medium; however
the left hand side is still evaluated in the heavy quark rest frame.

To rewrite $dp/dt$ on the left of \eq{momgain} in terms of the momentum
and time in the medium rest frame, $dp'/dt'$, note that $dp/dt$ is the 3-vector
component of the force 4-vector in the quark rest frame:
\be
f^x \equiv \frac{dp}{d\tau} = \frac{dp}{dt}.
\ee
One can see from \eq{pi1} that $\pi^1_t = 0$, and hence $f^t = 0$, 
and the 4-force boosted into the shock rest frame is therefore 
\be
f'^x = -\gamma f^x = -\gamma \frac{dp}{dt},
\ee
where the negative sign comes from boosting into a frame moving in 
the opposite direction; see \fig{frames}.  From the definition of the 
4-force we also know that in this frame
\be
f'^x \equiv \frac{dp'}{d\tau} = \gamma \frac{dp'}{dt'}.
\ee
Hence we find that
\be\label{dpprime}
\frac{dp}{dt} = -\frac{dp'}{dt'}.
\ee

Eqs.\ (\ref{momgain}), (\ref{mu}), and (\ref{dpprime}) lead us to the 
main result of this Letter:
\begin{align}
  \label{momgaintwo}
  \frac{d p'}{d t'} \, = \, - \frac{\sqrt{\lambda}}{2} \,
  \frac{\Lambda^2}{M_q} \, p'.
\end{align}
Should we take
the typical medium particle momentum to be $\Lambda = \sqrt{\pi}T$,
reminiscent of the thermal equilibrium result, then our result exactly
reproduces that of the black hole metric, $d p' / d t'
= - \pi \sqrt{\lambda} \, T^2 \, p'/ (2 \, M_q)$
\cite{Herzog:2006se,Gubser:2006bz,CasalderreySolana:2007qw}. 


\section{Limits of the Calculation}
\label{sec:limits}

As shown in
\cite{Herzog:2006gh,Gubser:2006bz,CasalderreySolana:2007qw} there are
limits to the applicability of the heavy quark drag calculations in a
black hole metric.  In the original analytic construction the heavy
quark was taken to have constant velocity, despite the momentum lost
to the medium.  If one proposes the existence of a field on the D7
brane supporting the finite mass heavy quark motion, then one is
naturally led to a speed limit based on the maximum field strength
supported by the Born-Infeld action.  However numerical calculations
were made for quarks with non-constant velocity and it turned out that
the drag form derived analytically is a very good approximation for
this more difficult problem.  Nevertheless one might still have a
speed limit induced by the warping of spacetime due to the presence of
the thermal black hole.  The finite heavy quark mass means that the
string endpoint representing the quark in 4D propagates at $z=z_M$ in
the $5^\mathrm{th}$ dimension (see \eq{zmass}). While at the Minkowski
boundary, $z=0$, the flatness of spacetime gives a local speed of
light $c=1$, reality of the point particle action in 5 bulk dimensions
in the AdS BH metric with the horizon at $z=z_h$ requires that
$\sqrt{-G_{\mu\nu} \, (dx^\mu/d\tau) \,  (dx^\nu/d\tau)} =
\sqrt{(1-(z/z_h)^4)/z^2-v^2/z^2}$ be real.  Then $v^2<1-(z/z_h)^4$
leading to $\gamma< (z_h/z)^2 < (z_h/z_M)^2$
\cite{Herzog:2006gh,Gubser:2006bz,CasalderreySolana:2007qw}.  The
speed limits resulting from these two lines of reasoning 
are precisely the same; this is
not an accident but due to T-duality \cite{Bachas:1995kx}.

While the metric, \eq{metric}, does not support an event horizon,
reality of the point particle action still yields an asymmetric,
$z$-dependent result for the local speed of light; to wit, in the rest
frame of the heavy quark,
\begin{align}
  \label{sl}
  \frac{\mu z^4-1}{\mu z^4 + 1} \le v \le 1.
\end{align}
At the boundary, $z=0$, the usual Minkowski speed of light limit,
$-1\le v\le 1$, is recovered. Motion at the stack of D3 color branes
(at $z=\infty$) is restricted to the speed of light in the direction
of the shock medium motion.  

In a similar way one finds that reality
of the Nambu-Goto action requires the velocity of the string at
$z=\infinity$ to be $+1$. This provides an argument against the 
reversed trailing string solution (given by \eq{zcubedsoln} with
the minus sign on its right hand side): the only classical fluctuations with non-zero velocity supported by
it are those that, at the D3
branes at $z=\infinity$, give motion at the speed of light in the
direction of the shock motion, i.e., towards the physical trailing
string solution (given by \eq{zcubedsoln} with the plus sign on the
right hand side).  This demonstrates that the reversed trailing string
solution is unstable under a large set of string fluctuations, and can
therefore be discarded. (This conclusion should be compared to the
stability analysis carried out in \cite{Gubser:2006nz} for the black
hole metric.)

Plugging $v=0$ for our static quark into \eq{sl} gives us the bound
for this calculation, namely
\begin{align}
\label{zmlimit}
  \mu \, z_M^4 \le 1.
\end{align}
Using \eq{zmass} to write $z_M = \sqrt{\lambda}/(2\pi M_q)$ and \eq{mu}
for $\mu=\pi^2\Lambda^4\gamma^2$, along with $\Lambda = \sqrt{\pi} \,
T$, we obtain
\begin{align}
  \gamma \le \frac{4 \, \pi \, M_q^2}{\lambda \, \Lambda^2} = \frac{4
    \, M_q^2}{\lambda \, T^2}.
\end{align}
The speed limit in this geometry is therefore identical to
that for the BH metric \cite{Herzog:2006gh,Gubser:2006nz,CasalderreySolana:2007qw}. 


\section{Conclusions and Discussions}
\label{sec:conc}

In this Letter we used the AdS/CFT correspondence to calculate the
drag force on a heavy quark in a strongly coupled medium.  By placing
the string in a shock metric, we extended the derivation of drag in a
thermalized medium---whose metric contains a black hole---to a
generalized medium not necessarily thermalized.  It turns out that,
just as for perturbative energy loss, the form of the drag is
independent of whether the medium is thermal or not.  Moreover the
momentum speed limit for the applicability of our calculations in the
shock metric is the same as was found for the black hole metric.

To complete the comparison of our calculation with the existing
ones for the drag force in a thermal medium
\cite{Herzog:2006gh,Gubser:2006bz,CasalderreySolana:2007qw} let us
find the mass of the heavy quark in the shock wave. According to the
standard prescription
\cite{Herzog:2006gh,Gubser:2006bz,CasalderreySolana:2007qw} the
(static) quark mass in the shock is
\begin{align}\label{mass}
  M_q^{shock} \, = \, - \int\limits_{z_M}^\infty dz \, \pi^0_t
\end{align}
which, with the help of \eq{pi0} and \eq{zcubedsoln} yields
\begin{align}\label{mass2}
  M_q^{shock} \, = \, \frac{\sqrt{\lambda}}{2 \, \pi} \,
  \int\limits_{z_M}^\infty \frac{dz}{z^2} \, = \,
  \frac{\sqrt{\lambda}}{2 \, \pi} \, \frac{1}{z_M} \, = \, M_q.
\end{align}
We see that the quark mass does not change inside the shock wave as
compared to the mass in the empty space outside of the shock wave.
This is in stark contrast to the thermal modification of the quark
mass observed in \cite{Herzog:2006gh} for a quark in a
finite-temperature medium.

To understand this result let us first discuss the origin of the
medium modification of the quark mass found in \cite{Herzog:2006gh}.
\footnote{We thank Cyrille Marquet for a discussion which led us to
  the qualitative understanding of the medium modification mechanism
  described below.} In \cite{Herzog:2006gh} it was found that at
finite temperature $T$ the quark mass becomes $M_q (T) =
(\sqrt{\lambda}/ 2 \pi) \, [(1/z_M) - (1/z_h)]$ with the horizon
radius $z_h = 1/\pi T$.  The mass of the heavy quark in AdS/CFT is
constructed by integration over the length of the string, as shown in
\eq{mass}. On the gauge theory side this implies that the mass is
built up from radiative corrections: at strong coupling the
corrections are large and lead to a large dressed quark mass
\cite{Dominguez:2008vd}. That is why the quarks in AdS/CFT are always
heavy. Thinking of a heavy quark as being dressed by a cloud of
strongly coupled particles (``partons'') it is easy to understand the
field theoretic origin of medium modification of the quark's mass, due 
on the AdS side 
to the $z_h$ IR cutoff in the $z$-integration introduced by the BH 
horizon \cite{Chernicoff:2008sa,Beuf:2008ep}:
the medium screens the infrared part of the parton cloud surrounding
the quark, thus reducing the dressed heavy quark mass. It is essential
that the quark is heavy not due to a large bare quark mass, but due to
quantum dressing by the gauge field fluctuations: this allows the
thermal medium to eliminate the soft IR fluctuations and make the
quark mass smaller in the medium.

Now, if we think of the shock wave considered here as of a boosted
thermal medium (which is one of the possibilities), then the time
scale of the thermal motion of the medium particles would be
Lorentz-dilated in the rest frame of the quark: $\tau_{th} \sim (1/T)
\, \gamma$. As $\gamma \gg 1$, the dilated thermal time scale
$\tau_{th}$ is much longer than any other time scale in the problem.
This is a well-known feature of the parton model in the infinite
momentum frame, in which the partons appear ``frozen in time'' to the
probe interacting with them.  Hence our static heavy quark does not
``feel'' the thermal motion of the particles in the shock wave over
the typical interaction time scale between the quark and a shock wave
particle. This is why the heavy quark mass does not get modified by
the shock wave, as we saw in \eq{mass2}: the lack of thermal motion on
the interaction time scale means no screening for the infrared partons
in the quark wave function. Indeed there is an IR screening due to
saturation effects in the shock
\cite{Albacete:2008ze,Dominguez:2008vd}, which may lead to quark mass
modification if we integrate over $z$ in \eq{mass2} only up to the
horizon of the induced metric on the string, as was done in
\cite{Dominguez:2008vd,Beuf:2008ep}. Our qualitative picture also explains why the
thermal effects are not relevant for the quark energy loss: in the
infinite momentum frame discussed here the thermal motion is
irrelevant, but the quark still experiences a drag force. We conclude
that the drag force indeed does not depend on the thermal motion, and
is thus independent of whether the medium is thermal or not. This is
exactly the same conclusion as we obtained by explicit derivation
above.


\section{Acknowledgments}

We are grateful to Alberto Guijosa, Miklos Gyulassy, Ulrich Heinz, Boris Kopeliovich,
Hong Liu, Cyrille Marquet, Samir Mathur, Jorge Noronha, Krishna
Rajagopal, Anastasios Taliotis, Derek Teaney, and Bill Zajc for fruitful
discussions.  

This work is supported by the Office of Nuclear Physics, of the Office
of Science, of the U.S.\ Department of Energy under Grant No.\ 
DE-FG02-05ER41377.


\appendix
\renewcommand{\theequation}{A.\arabic{equation}}
\setcounter{equation}{0}
\section{Other Solutions to the EOM}\label{sec:othersolns}

The argument used to derive the stable static solutions $x(t,z)=x_0$,
$x_0\pm \sqrt{\mu}z^3/3$ required the string to hang from a $z$
smaller than both $\sqrt[4]{\mu}$ and $\sqrt{C}$ down to the stack of
D3 branes at $z=\infinity$ so that, in order for the solution to
remain real, both the numerator and denominator of the radical in
\eq{xiprime} had to change sign at precisely the same value of $z$.
However strings with a turning point may also satisfy this reality
condition.

For such a solution $x(z)$ to exist, $z(x)$ should increase with $x$
until the turning point, where the slope of $z(x)$ must be 0 (note the
use of the inverse function).  If we denote the value of $z$ for which
the numerator of the radical in \eq{xiprime} is zero as $z_c$ and the
value of $z$ for which the denominator is zero as $z_{max}$, then
\begin{equation}
\begin{array}{lccc}
& 1-\mu z_c^4 = 0 & \qquad & 1-C^2 z_{max}^4 = 0 \\
\Rightarrow & z_c = \mu^{-1/4} & \qquad & z_{max} = C^{-1/2}.
\end{array}
\end{equation}
The condition for turning, $dz/dx = 0$, thus requires the string
solution to extend up to $z=z_{max}$.

Concentrating on stable string solutions, ones for which the string
configuration is purely real, we can consider two cases: (1) $z_M <
z_{max} < z_c$ and (2) $z_M > z_{max}$, $z_c$.  The latter case may be
solved analytically in closed form for the two shape possibilities:
(a) string endpoints at plus and minus infinity in the $x^3$
direction, both terminating on the D7 or the stack of D3 branes with a
hump at $x_0$ that passes through $z_{max}$ and (b) a hanging string
solution from the D7 brane to the D3 branes.  There is considerable
freedom in these solutions as $C$ is no longer fixed.  However none
are physical due to the speed limit condition, \eq{zmlimit}.

\begin{figure}[!htb]
\centering
\includegraphics[width=12cm]{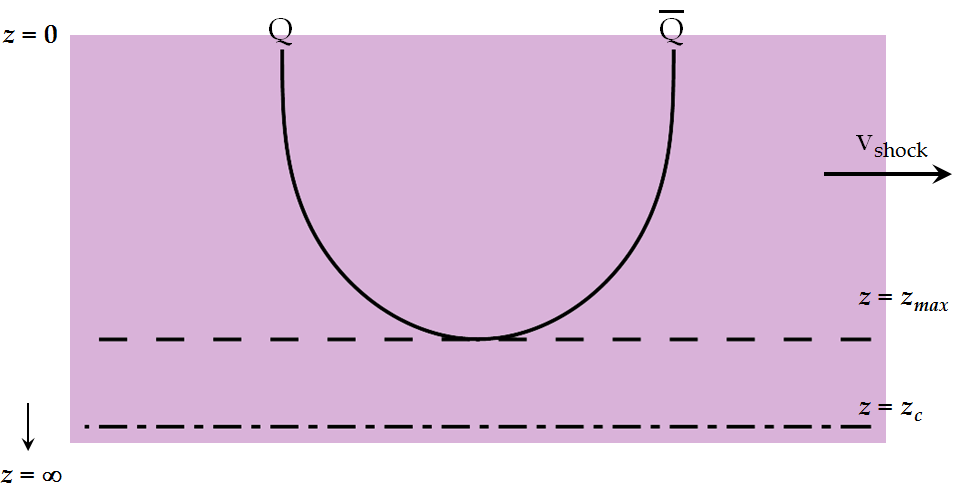}
\caption{Visualization of the hanging meson configuration, \eq{appellone}.  Note that $z_c\,>\,z_{max}$, and the string turns precisely at $z=z_{max}$.}
\label{appellonesoln}
\end{figure}

Unlike in the black hole metric one may also solve the hanging
``meson'' case (1) analytically.  For simplicity let us consider
infinitely massive quarks with $z_M=0$.  The parameterization from
$x=-r/2$ to 0 for a string with ends at $(x,z) = (\pm r/2, 0)$ may be
found by direct integration of the first integral of the motion,
\eq{xiprime}:
\begin{align}
  \xi(z) & = -\frac{r}{2} + \frac{1}{z_{max}^2} \, \int\limits_0^{z} 
dz' z'^2 \sqrt{ \frac{1-\mu z'^4}{1 - \frac{z'^4}{z_{max}^4}} } \nonumber\\
\label{appellone}
& = -\frac{r}{2} + \frac{z^3}{3 \, z_{max}^2} \, \mathrm{F}_1
\left(\frac{3}{4};-\frac{1}{2},\frac{1}{2};\frac{7}{4};\mu \, z^4,
  \frac{z^4}{z_{max}^4}\right),
\end{align}
where $\mathrm{F}_1$ is Appell's hypergeometric function of the first
kind (to construct the solution in the interval from $x=0$ to $r/2$
simply reflect \eq{appellone} across $x=0$); this solution is shown in \fig{appellonesoln}.  The solution in
\eq{appellone} complements the results of \cite{Albacete:2008ze} in
which the string profile was found for a dipole oriented perpendicular
to the direction of shock wave motion, by providing the solution for
a dipole oriented parallel to the shock velocity.

Imposing the turning condition on \eq{appellone}, i.e., that in the
middle of the string at $x=0$ one has the turning point $z=z_{max}$,
yields the following condition for $z_{max}$:
\begin{align}
  \frac{r}{2} \, = \, \frac{\sqrt{\pi} \, \Gamma (7/4) }{3 \, \Gamma
    (5/4)} \, z_{max} \, \mathrm{F} \left( \frac{3}{4}, - \frac{1}{2};
    \frac{5}{4}; \mu \, z_{max}^4 \right).
\end{align}
Numerical analysis shows that for $r > r_{crit} \approx 0.85
\, z_c = 0.85 \, \mu^{-1/4}$ the solution for $z_{max}$ becomes
complex-valued.

For $r<r_{crit}$ the singlet configuration for the quark anti-quark
pair, \eq{appellone}, suffers no energy loss: the momentum lost by the
leading heavy quark is exactly canceled by the momentum gain of the
heavy quark that follows behind. However, for separations $r >
r_{crit}$ the system is unstable to decay into a color-adjoint state,
possibly leading to complete dissolution of the meson. To completely
answer this question a full study of the roots for $z_{max}$ in the
case of a general orientation of the dipole with respect to the shock
velocity similar to what was done in \cite{Liu:2006he} needs to be
carried out: such analysis goes beyond the scope of this work.


\phantomsection


\addcontentsline{toc}{chapter}{References}

\begin{thebibliography}{10}
\providecommand{\url}[1]{#1}
\providecommand{\urlprefix}{URL }
\providecommand{\eprint}[2][]{[arXiv:\url{#2}]}

\bibitem{Maldacena:1997re}
J.~M. Maldacena, \emph{{The large N limit of superconformal field theories and
  supergravity}}, Adv. Theor. Math. Phys. \textbf{2}, 231 (1998)
  \eprint{hep-th/9711200}.

\bibitem{Gubser:1998bc}
S.~S. Gubser, I.~R. Klebanov, and A.~M. Polyakov, \emph{{Gauge theory
  correlators from non-critical string theory}}, Phys. Lett. \textbf{B428}, 105
  (1998) \eprint{hep-th/9802109}.

\bibitem{Witten:1998qj}
E.~Witten, \emph{{Anti-de Sitter space and holography}}, Adv. Theor. Math.
  Phys. \textbf{2}, 253 (1998) \eprint{hep-th/9802150}.

\bibitem{Aharony:1999ti}
O.~Aharony, S.~S. Gubser, J.~M. Maldacena, H.~Ooguri, and Y.~Oz, \emph{{Large N
  field theories, string theory and gravity}}, Phys. Rept. \textbf{323}, 183
  (2000) \eprint{hep-th/9905111}.

\bibitem{Gubser:2009md}
S.~S. Gubser and A.~Karch, \emph{{From gauge-string duality to strong
  interactions: a Pedestrian's Guide}}  (2009) \eprint{0901.0935}.

\bibitem{Gubser:2009sn}
S.~S. Gubser, S.~S. Pufu, F.~D. Rocha, and A.~Yarom, \emph{{Energy loss in a
  strongly coupled thermal medium and the gauge-string duality}}  (2009)
  \eprint{0902.4041}.

\bibitem{Wicks:2005gt}
S.~Wicks, W.~Horowitz, M.~Djordjevic, and M.~Gyulassy, \emph{{Elastic,
  Inelastic, and Path Length Fluctuations in Jet Tomography}}, Nucl. Phys.
  \textbf{A784}, 426 (2007) \eprint{nucl-th/0512076}.

\bibitem{Kovtun:2004de}
P.~Kovtun, D.~T. Son, and A.~O. Starinets, \emph{{Viscosity in strongly
  interacting quantum field theories from black hole physics}}, Phys. Rev.
  Lett. \textbf{94}, 111601 (2005) \eprint{hep-th/0405231}.

\bibitem{Janik:2005zt}
R.~A. Janik and R.~B. Peschanski, \emph{{Asymptotic perfect fluid dynamics as a
  consequence of AdS/CFT}}, Phys. Rev. \textbf{D73}, 045013 (2006)
  \eprint{hep-th/0512162}.

\bibitem{Liu:2006nn}
H.~Liu, K.~Rajagopal, and U.~A. Wiedemann, \emph{{An AdS/CFT calculation of
  screening in a hot wind}}, Phys. Rev. Lett. \textbf{98}, 182301 (2007)
  \eprint{hep-ph/0607062}.

\bibitem{Herzog:2006se}
C.~P. Herzog, \emph{{Energy loss of heavy quarks from asymptotically AdS
  geometries}}, JHEP \textbf{09}, 032 (2006) \eprint{hep-th/0605191}.

\bibitem{Gubser:2006bz}
S.~S. Gubser, \emph{{Drag force in AdS/CFT}}, Phys. Rev. \textbf{D74}, 126005
  (2006) \eprint{hep-th/0605182}.

\bibitem{CasalderreySolana:2007qw}
J.~Casalderrey-Solana and D.~Teaney, \emph{{Transverse momentum broadening of a
  fast quark in a N = 4 Yang Mills plasma}}, JHEP \textbf{04}, 039 (2007)
  \eprint{hep-th/0701123}.

\bibitem{Horowitz:2007su}
W.~A. Horowitz and M.~Gyulassy, \emph{{Heavy quark jet tomography of Pb + Pb at
  LHC: AdS/CFT drag or pQCD energy loss?}}, Phys. Lett. \textbf{B666}, 320
  (2008) \eprint{0706.2336}.

\bibitem{Mikhailov:2003er}
A.~Mikhailov, \emph{{Nonlinear waves in AdS/CFT correspondence}}  (2003)
  \eprint{hep-th/0305196}.

\bibitem{Sin:2004yx}
S.-J. Sin and I.~Zahed, \emph{{Holography of radiation and jet quenching}},
  Phys. Lett. \textbf{B608}, 265 (2005) \eprint{hep-th/0407215}.

\bibitem{Chernicoff:2008sa}
M.~Chernicoff and A.~Guijosa, \emph{{Acceleration, Energy Loss and Screening in
  Strongly- Coupled Gauge Theories}}, JHEP \textbf{06}, 005 (2008)
  \eprint{0803.3070}.

\bibitem{Kharzeev:2008qr}
D.~E. Kharzeev, \emph{{Universal upper bound on the energy of a parton escaping
  from the strongly coupled quark-gluon matter}}  (2008) \eprint{0806.0358}.

\bibitem{Chesler:2007sv}
P.~M. Chesler and L.~G. Yaffe, \emph{{The stress-energy tensor of a quark
  moving through a strongly-coupled N=4 supersymmetric Yang-Mills plasma:
  comparing hydrodynamics and AdS/CFT}}, Phys. Rev. \textbf{D78}, 045013 (2008)
  \eprint{0712.0050}.

\bibitem{Karsch:2000ps}
F.~Karsch, E.~Laermann, and A.~Peikert, \emph{{The pressure in 2, 2+1 and 3
  flavour QCD}}, Phys. Lett. \textbf{B478}, 447 (2000)
  \eprint{hep-lat/0002003}.

\bibitem{Cheng:2007jq}
M.~Cheng et~al., \emph{{The QCD Equation of State with almost Physical Quark
  Masses}}, Phys. Rev. \textbf{D77}, 014511 (2008) \eprint{0710.0354}.

\bibitem{Gubser:1996de}
S.~S. Gubser, I.~R. Klebanov, and A.~W. Peet, \emph{{Entropy and Temperature of
  Black 3-Branes}}, Phys. Rev. \textbf{D54}, 3915 (1996)
  \eprint{hep-th/9602135}.

\bibitem{Klebanov:2000me}
I.~R. Klebanov, \emph{{TASI lectures: Introduction to the AdS/CFT
  correspondence}}  (2000) \eprint{hep-th/0009139}.

\bibitem{Teaney:1999gr}
D.~Teaney and E.~V. Shuryak, \emph{{An unusual space-time evolution for heavy
  ion collisions at high energies due to the QCD phase transition}}, Phys. Rev.
  Lett. \textbf{83}, 4951 (1999) \eprint{nucl-th/9904006}.

\bibitem{Kolb:2000sd}
P.~F. Kolb, J.~Sollfrank, and U.~W. Heinz, \emph{{Anisotropic transverse flow
  and the quark-hadron phase transition}}, Phys. Rev. \textbf{C62}, 054909
  (2000) \eprint{hep-ph/0006129}.

\bibitem{Huovinen:2001cy}
P.~Huovinen, P.~F. Kolb, U.~W. Heinz, P.~V. Ruuskanen, and S.~A. Voloshin,
  \emph{{Radial and elliptic flow at RHIC: further predictions}}, Phys. Lett.
  \textbf{B503}, 58 (2001) \eprint{hep-ph/0101136}.

\bibitem{Teaney:2003kp}
D.~Teaney, \emph{{Effect of shear viscosity on spectra, elliptic flow, and
  Hanbury Brown-Twiss radii}}, Phys. Rev. \textbf{C68}, 034913 (2003)
  \eprint{nucl-th/0301099}.

\bibitem{Song:2007ux}
H.~Song and U.~W. Heinz, \emph{{Causal viscous hydrodynamics in 2+1 dimensions
  for relativistic heavy-ion collisions}}, Phys. Rev. \textbf{C77}, 064901
  (2008) \eprint{0712.3715}.

\bibitem{Adler:2005xv}
S.~S. Adler et~al. (PHENIX), \emph{{Nuclear modification of electron spectra
  and implications for heavy quark energy loss in Au + Au collisions at
  s(NN)**(1/2) = 200-GeV}}, Phys. Rev. Lett. \textbf{96}, 032301 (2006)
  \eprint{nucl-ex/0510047}.

\bibitem{Abelev:2006db}
B.~I. Abelev et~al. (STAR), \emph{{Transverse momentum and centrality
  dependence of high-pt non-photonic electron suppression in Au+Au collisions
  at $\sqrt{s_{NN}}$ = 200 GeV}}, Phys. Rev. Lett. \textbf{98}, 192301 (2007)
  \eprint{nucl-ex/0607012}.

\bibitem{Adare:2006nq}
A.~Adare et~al. (PHENIX), \emph{{Energy Loss and Flow of Heavy Quarks in Au+Au
  Collisions at $\sqrt{s_NN}$ = 200 GeV}}, Phys. Rev. Lett. \textbf{98}, 172301
  (2007) \eprint{nucl-ex/0611018}.

\bibitem{Mischke:2008qj}
A.~Mischke (STAR), \emph{{Heavy-flavor particle correlations in STAR via
  electron azimuthal correlations with $D^0$ mesons}}, J. Phys. \textbf{G35},
  104117 (2008) \eprint{0804.4601}.

\bibitem{Morino:2008nc}
Y.~Morino (PHENIX), \emph{{Measurement of charm and bottom production in p+p
  collisions at $\sqrt{s}$ = 200 GeV at RHIC-PHENIX}}, J. Phys. \textbf{G35},
  104116 (2008) \eprint{0805.3871}.

\bibitem{Adams:2005ph}
J.~Adams et~al. (STAR), \emph{{Distributions of charged hadrons associated with
  high transverse momentum particles in p p and Au + Au collisions at
  s(NN)**(1/2) = 200-GeV}}, Phys. Rev. Lett. \textbf{95}, 152301 (2005)
  \eprint{nucl-ex/0501016}.

\bibitem{Adler:2005ee}
S.~S. Adler et~al. (PHENIX), \emph{{Modifications to di-jet hadron pair
  correlations in Au + Au collisions at s(NN)**(1/2) = 200-GeV}}, Phys. Rev.
  Lett. \textbf{97}, 052301 (2006) \eprint{nucl-ex/0507004}.

\bibitem{Adare:2008cqb}
A.~Adare et~al. (PHENIX), \emph{{Dihadron azimuthal correlations in Au+Au
  collisions at $\sqrt{s_{NN}}$=200 GeV}}, Phys. Rev. \textbf{C78}, 014901
  (2008) \eprint{0801.4545}.

\bibitem{Ulery:2005cc}
J.~G. Ulery (STAR), \emph{{Two- and three-particle jet correlations from
  STAR}}, Nucl. Phys. \textbf{A774}, 581 (2006) \eprint{nucl-ex/0510055}.

\bibitem{Zhang:2007zzp}
C.~Zhang (PHENIX), \emph{{Studying the medium modification of jets via
  high-p(T) hadron angular correlations}}, J. Phys. \textbf{G34}, S671 (2007).

\bibitem{Stoecker:2004qu}
H.~Stoecker, \emph{{Collective Flow signals the Quark Gluon Plasma}}, Nucl.
  Phys. \textbf{A750}, 121 (2005) \eprint{nucl-th/0406018}.

\bibitem{CasalderreySolana:2004qm}
J.~Casalderrey-Solana, E.~V. Shuryak, and D.~Teaney, \emph{{Conical flow
  induced by quenched QCD jets}}, J. Phys. Conf. Ser. \textbf{27}, 22 (2005)
  \eprint{hep-ph/0411315}.

\bibitem{Friess:2006fk}
J.~J. Friess, S.~S. Gubser, G.~Michalogiorgakis, and S.~S. Pufu, \emph{{The
  stress tensor of a quark moving through N = 4 thermal plasma}}, Phys. Rev.
  \textbf{D75}, 106003 (2007) \eprint{hep-th/0607022}.

\bibitem{Betz:2008wy}
B.~Betz, M.~Gyulassy, J.~Noronha, and G.~Torrieri, \emph{{Anomalous Conical
  Di-jet Correlations in pQCD vs AdS/CFT}}  (2008) \eprint{0807.4526}.

\bibitem{Baier:1996kr}
R.~Baier, Y.~L. Dokshitzer, A.~H. Mueller, S.~Peigne, and D.~Schiff,
  \emph{{Radiative energy loss of high energy quarks and gluons in a
  finite-volume quark-gluon plasma}}, Nucl. Phys. \textbf{B483}, 291 (1997)
  \eprint{hep-ph/9607355}.

\bibitem{Gyulassy:2000er}
M.~Gyulassy, P.~Levai, and I.~Vitev, \emph{{Reaction operator approach to
  non-Abelian energy loss}}, Nucl. Phys. \textbf{B594}, 371 (2001)
  \eprint{nucl-th/0006010}.

\bibitem{Vitev:2002pf}
I.~Vitev and M.~Gyulassy, \emph{{High $p_{T}$ tomography of $d$ + Au and Au+Au
  at SPS, RHIC, and LHC}}, Phys. Rev. Lett. \textbf{89}, 252301 (2002)
  \eprint{hep-ph/0209161}.

\bibitem{Adare:2008qa}
A.~Adare et~al. (PHENIX), \emph{{Suppression pattern of neutral pions at high
  transverse momentum in Au+Au collisions at $\sqrt{s_NN}$ = 200 GeV and
  constraints on medium transport coefficients}}, Phys. Rev. Lett.
  \textbf{101}, 232301 (2008) \eprint{0801.4020}.

\bibitem{Lin:2008zzi}
G.~Lin (STAR), \emph{{First STAR results on pi0 production over an extended
  p(T)-range from 200-GeV Au + Au collisions}}, J. Phys. \textbf{G35}, 104046
  (2008).

\bibitem{Adler:2006bv}
S.~S. Adler et~al. (PHENIX), \emph{{High transverse momentum $\eta$ meson
  production in $p^+ p$, $d^+$ Au and Au+Au collisions at $S(NN) ^{(1/2)}$ =
  200-GeV}}, Phys. Rev. \textbf{C75}, 024909 (2007) \eprint{nucl-ex/0611006}.

\bibitem{Adler:2005ig}
S.~S. Adler et~al. (PHENIX), \emph{{Centrality dependence of direct photon
  production in s(NN)**(1/2) = 200-GeV Au + Au collisions}}, Phys. Rev. Lett.
  \textbf{94}, 232301 (2005) \eprint{nucl-ex/0503003}.

\bibitem{Miki:2008zz}
K.~Miki (PHENIX), \emph{{High-p(T) direct photon spectra and azimuthal
  anisotropy measurements in 200-GeV Au + Au collisions at RHIC- PHENIX}}, J.
  Phys. \textbf{G35}, 104122 (2008).

\bibitem{Djordjevic:2005db}
M.~Djordjevic, M.~Gyulassy, R.~Vogt, and S.~Wicks, \emph{{Influence of bottom
  quark jet quenching on single electron tomography of Au + Au}}, Phys. Lett.
  \textbf{B632}, 81 (2006) \eprint{nucl-th/0507019}.

\bibitem{Armesto:2005mz}
N.~Armesto, M.~Cacciari, A.~Dainese, C.~A. Salgado, and U.~A. Wiedemann,
  \emph{{How sensitive are high-p(T) electron spectra at RHIC to heavy quark
  energy loss?}}, Phys. Lett. \textbf{B637}, 362 (2006)
  \eprint{hep-ph/0511257}.

\bibitem{Adler:2006bw}
S.~S. Adler et~al. (PHENIX), \emph{{A detailed study of high-p(T) neutral pion
  suppression and azimuthal anisotropy in Au + Au collisions at s(NN)**(1/2) =
  200-GeV}}, Phys. Rev. \textbf{C76}, 034904 (2007) \eprint{nucl-ex/0611007}.

\bibitem{Abelev:2008ed}
B.~I. Abelev et~al. (STAR), \emph{{Centrality dependence of charged hadron and
  strange hadron elliptic flow from $\sqrt{s_{NN}}$ = 200 GeV Au+Au
  collisions}}, Phys. Rev. \textbf{C77}, 054901 (2008) \eprint{0801.3466}.

\bibitem{Gyulassy:1993hr}
M.~Gyulassy and X.-n. Wang, \emph{{Multiple collisions and induced gluon
  Bremsstrahlung in QCD}}, Nucl. Phys. \textbf{B420}, 583 (1994)
  \eprint{nucl-th/9306003}.

\bibitem{Collins:1992xw}
J.~C. Collins, \emph{{Hard scattering in QCD with polarized beams}}, Nucl.
  Phys. \textbf{B394}, 169 (1993) \eprint{hep-ph/9207265}.

\bibitem{Collins:2007nk}
J.~Collins and J.-W. Qiu, \emph{{$k_{T}$ factorization is violated in
  production of high- transverse-momentum particles in hadron-hadron
  collisions}}, Phys. Rev. \textbf{D75}, 114014 (2007) \eprint{0705.2141}.

\bibitem{deHaro:2000xn}
S.~de~Haro, S.~N. Solodukhin, and K.~Skenderis, \emph{{Holographic
  reconstruction of spacetime and renormalization in the AdS/CFT
  correspondence}}, Commun. Math. Phys. \textbf{217}, 595 (2001)
  \eprint{hep-th/0002230}.

\bibitem{Fefferman}
C.~Fefferman and C.~R. Graham, \emph{Conformal Invariants}, in \emph{\'Elie
  Cartan et les Math\'ematiques d'aujourd'hui}, 95--116 (Ast\'erisque, 1985).

\bibitem{Albacete:2008ze}
J.~L. Albacete, Y.~V. Kovchegov, and A.~Taliotis, \emph{{DIS on a Large Nucleus
  in AdS/CFT}}, JHEP \textbf{07}, 074 (2008) \eprint{0806.1484}.

\bibitem{Albacete:2008vs}
J.~L. Albacete, Y.~V. Kovchegov, and A.~Taliotis, \emph{{Modeling Heavy Ion
  Collisions in AdS/CFT}}, JHEP \textbf{07}, 100 (2008) \eprint{0805.2927}.

\bibitem{Karch:2002sh}
A.~Karch and E.~Katz, \emph{{Adding flavor to AdS/CFT}}, JHEP \textbf{06}, 043
  (2002) \eprint{hep-th/0205236}.

\bibitem{Herzog:2006gh}
C.~P. Herzog, A.~Karch, P.~Kovtun, C.~Kozcaz, and L.~G. Yaffe, \emph{{Energy
  loss of a heavy quark moving through N = 4 supersymmetric Yang-Mills
  plasma}}, JHEP \textbf{07}, 013 (2006) \eprint{hep-th/0605158}.

\bibitem{Bachas:1995kx}
C.~Bachas, \emph{{D-brane dynamics}}, Phys. Lett. \textbf{B374}, 37 (1996)
  \eprint{hep-th/9511043}.

\bibitem{Gubser:2006nz}
S.~S. Gubser, \emph{{Momentum fluctuations of heavy quarks in the gauge-string
  duality}}, Nucl. Phys. \textbf{B790}, 175 (2008) \eprint{hep-th/0612143}.

\bibitem{Dominguez:2008vd}
F.~Dominguez, C.~Marquet, A.~H. Mueller, B.~Wu, and B.-W. Xiao,
  \emph{{Comparing energy loss and $p_{\perp}$-broadening in perturbative QCD
  with strong coupling $\mathcal{N}=4$ SYM theory}}, Nucl. Phys. \textbf{A811},
  197 (2008) \eprint{0803.3234}.

\bibitem{Beuf:2008ep}
G.~Beuf, C.~Marquet, and B.-W. Xiao, \emph{{Heavy-quark energy loss and
  thermalization in a strongly coupled SYM plasma}}  (2008) \eprint{0812.1051}.

\bibitem{Liu:2006he}
H.~Liu, K.~Rajagopal, and U.~A. Wiedemann, \emph{{Wilson loops in heavy ion
  collisions and their calculation in AdS/CFT}}, JHEP \textbf{03}, 066 (2007)
  \eprint{hep-ph/0612168}.

\end{thebibliography}
\end{document}